# Breakdown of the Jarzynski relation for an adiabatic stretching of an isotropic spring

Jaeyoung Sung

*Department of Chemistry, Chung-Ang University, Seoul 156-756 Korea*

**Abstract**

We perform a theoretical test of Jarzynski relation for an adiabatic stretching of an isotropic spring, which is an exactly solvable model. It turns out that Jarzynski relation does not hold even when the entire infinite momentum space of the system state is taken into account, in contradiction to recent arguments supporting Jarzynski relation.

Jarzynski relation has drawn much recent attentions as it suggests a novel method for estimation of free energy change.[1-3] According to Jarzynski relation, the free energy difference $\Delta F$ between two states of a system is related to distribution of the amount $W$ of work done on the system during an arbitrary process connecting the two states of the system:

$$\exp(-\beta \Delta F) = \langle \exp(-\beta W) \rangle, \tag{1}$$

where $\beta$ is given by $(k_B T)^{-1}$ with $k_B$ and $T$ being the Boltzmann constant and the absolute temperature, respectively. Eq. (1) is first obtained for an adiabatic process in which a system evolves according to the classical dynamics,[1] and then generalized into other processes.[4-7] Given that the system is initially in thermal equilibrium with a heat bath, Eq. (1) has been believed to be valid irrespective of the nature of the process that induces the change of the system states,[2-9] as it is claimed in Ref. 1.

However, recently, Mauzurall and Cohen raised a question about the validity of Eq. (1) for processes that drive a system far from the thermal equilibrium state.[10] More recently, Sung derived the validity condition of Jarzynski relation for an adiabatic process for which dynamics of the system obeys classical mechanics.[11] According to Ref. 11, for an adiabatic process that changes a system coupled parameter $\lambda$ from $\lambda_0$ to $\lambda_1$ in time $t_S$, Jarzynski relation holds if and only if the phase space extension of the isolated system with $\lambda = \lambda_1$ at the very end of the adiabatic process, coincides with the phase space extension of the system with $\lambda = \lambda_1$ at thermal equilibrium. However, it is not yet well-known that the validity condition can be violated and Eq. (1) can break down when the equilibrium phase space extension of the initial state of the system is different from that of the final state.

For an example, let us consider the difference $\Delta F_{id}(v_a < v < v_b)$ of the free energy for $N$ number of ideal gas particles in a velocity interval ($v_a < v < v_b$) in volume $V_1$ at thermal equilibrium from that in volume $V_0$ at thermal equilibrium, which is given by $\Delta F_{id}(v_a < v < v_b) = -Nk_B T \ln(V_1/V_0)$ for any value for $v_a$ and $v_b$. However, $\Delta F_{id}(v_a < v < v_b)$ cannot estimated by Eq. (1) with adiabatic processes that changes the system volume from $V_0$ to $V_1$, unless $V_0$ is equal to $V_1$ for any value of $v_a$ and $v_b$ as the validity condition is not satisfied.[11] However, in the limiting case with $v_a \to -\infty$ and $v_b \to \infty$, the validity condition of Jarzynski relation turns out satisfied so that Eq. (1) yields correct expression for $\Delta F_{id}$ for the adiabatic expansion process in which the position of a boundary increases linearly in time.[11,12]

Based on the latter fact, it was asserted that Eq. (1) yields a correct result for the free energy change of a system state with the entire infinite momentum space extension. Although, for this case, it would not be feasible to use Eq. (1) in practice as one cannot sample the infinite momentum space to evaluate the R.H.S. of Eq. (1), it is an interesting question whether or not Eq. (1) is a correct equation if the entire infinite momentum space extension is taken into account.

A clue to the answer to the latter problem can be found in an adiabatic expansion into vacuum in which a boundary of the initial gas system is removed in the direction vertical to the direction of the gas expansion into vacuum. For the adiabatic expansion into vacuum, it can be shown that the validity condition of Eq. (1) is not satisfied, and Eq. (1) does not hold even if the entire infinite momentum space extension of the initial gas system is taken into account.[11,13] This example reflects the fact that the R.H.S. of Eq. (1) is not a state function in contrast to the L.H.S or the free energy function so that Eq. (1) does not hold in general even if the entire infinite phase space of an initial system is sampled completely according to the canonical distribution. Nevertheless, the latter fact is not yet well known, and the validity of Eq. (1) has been still asserted.[14] As for the reason for the failure of the Jarzynski relation to the ideal gas expansion into vacuum, it was argued that a gas system cannot be in canonical equilibrium before a vacuum expansion process, while it can before other expansion processes of gases. However, the argument is hardly justifiable as the canonical ensemble of a system at thermal equilibrium can be defined without anything to do with the process the system suffers after the equilibrium state.

In the present work, hoping to settle down this controversial issue, we will consider a system in an isotropic potential, for which an exact analysis can be done. Although the model is simple, it is enough to demonstrate the fact that Eq. (1) does not hold even if the initial phase space extension of the system is completely sampled according to canonical distribution, in contradiction to recent argument supporting the validity of Jarzynski relation.

The Hamiltonian of the system we consider is given by

$$H_\mathbf{r}(\mathbf{p},\mathbf{q}) = H_0(\mathbf{p},\mathbf{q}) + U(|\mathbf{r}|), \qquad (2)$$

where $H_0(\mathbf{p},\mathbf{q})$ denotes the part of the system in the absence of the potential $U$. The potential field $U$ is dependent on the magnitude $r[\equiv |\mathbf{r}|]$ of a vector $\mathbf{r}$, which is our control parameter. Although it is not necessary, for concreteness, we will choose $H_0(\mathbf{p},\mathbf{q})$ and $U(|\mathbf{r}|)$ to be the Hamiltonian of a rigid rotor and the potential of an isotropic spring, respectively. The explicit form of the rigid rotor Hamiltonian is given by

$$H_0(\mathbf{p},\mathbf{q}) = \frac{1}{2I}(p_\theta^2 + \frac{1}{\sin^2\theta} p_\phi^2), \qquad (3)$$

where $I$ denotes the moment of inertia. $p_\theta$ and $p_\phi$ are the momenta conjugated to the polar angle $\theta$ and the azimuthal angle $\psi$ specifying the direction of the rigid rotor in a spherical polar coordinate. **p** and **q** denote two-dimensional vectors defined by $\mathbf{p} = (p_\theta, p_\psi)$ and $\mathbf{q} = (\theta, \psi)$. The model system with the latter choice can be pictured as a rigid rotor located at the end of an isotropic spring whose position can be controlled parametrically.

If we choose $r$ and T as the thermodynamic state variables of our system, the classical partition function $Q(R,T)$ of the thermodynamic state $(r = R, T)$ of the system is given by

$$Q(R,T) = \int_{|\mathbf{r}|=R} d\mathbf{r} \int d\mathbf{p} \int d\mathbf{q} \exp[-\beta H_\mathbf{r}(\mathbf{p},\mathbf{q})]$$
$$= 4\pi R^2 \exp[-\beta U(R)](8\pi^2 I k_B T) \quad . \tag{4}$$

Therefore, Helmholtz free energy difference $\Delta F(T)[\equiv F(R_1,T) - F(R_0,T)]$ between thermodynamic states $(r = R_1, T)$ and $(r = R_0, T)$ is given by

$$\Delta F(T) = U(R_1) - U(R_0) - 2k_B T \ln(R_1/R_0) . \tag{5}$$

Since the thermodynamic internal energy $\overline{U}(r,T)$ of the system in state $(r,T)$ is given by $\overline{U}(r,T) = k_B T + U(r)$, $U(R_1) - U(R_0)$ in the R.H.S. of Eq. (5) is nothing but the difference $\Delta \overline{U}(T)[\equiv \overline{U}(R_1,T) - \overline{U}(R_0,T)]$ in the thermodynamic internal energy. From the definition of Helmholtz free energy, one obtains the entropy change as $\Delta \overline{S} = 2k_B \ln(R_1/R_0)$ from Eq. (5).

Now we calculate the free energy change from Jarzynski relation given in Eq. (1) and compare the result to Eq. (5). Let the system be initially in thermal equilibrium with a heat bath. At time zero, we isolate the system from the heat bath and initiate the adiabatic process in which we change the radius $r$ of the isotropic spring from $R_0$ to a greater value $R_1$. Let the adiabatic process be completed at time $t_S$, after which we get the system in touch with the heat bath and let the system relax to the thermal equilibrium state $(R_1, T)$. As we don't apply any force on the rigid rotor throughout the adiabatic stretching of the isotropic spring, the angular momentum and the rotational kinetic energy of the rotor conserves and the mechanical work $W$ done on the total system during the adiabatic spring stretching process is the same as the difference $U(R_1) - U(R_0)$ in the potential energy of the spring for every initial microscopic state in the thermodynamic state $(r = R_0, T)$ of the rotor. For the latter adiabatic process, evaluation of $\langle \exp(-\beta W) \rangle$ is trivial and $\Delta F^J$ calculated from Eq. (1) is given by

$$\Delta F^J = U(R_1) - U(R_0) . \tag{6}$$

From the comparison between Eq. (5) and Eq. (6), one can see that the Eq. (1) does not hold even though the initial phase space extension of the system is completely sampled according to canonical distribution in the evaluation of the R.H.S. of Eq. (1), which is in direct contradiction to recent argument supporting the validity of Jarzynski relation.

In the similar way, one can show that Eq. (1) does not hold for the adiabatic stretching of the system with a $d$-dimensional isotropic spring $(d \geq 2)$. For the latter system, the free energy change from state $(r = R_0, T)$ to $(r = R_1, T)$ becomes

$$\Delta F_d(T) = U(R_1) - U(R_0) - (d-1)k_B T \ln(R_1/R_0), \tag{7}$$

whereas the free energy change $\Delta F^J$ calculated from Jarzynski relation for the adiabatic stretching process is given by Eq. (6) irrespective of the spatial dimension $d$.

The model we have considered here and the adiabatic expansion of a gas into vacuum are examples of the cases where the validity condition of Jarzynski relation is violated. For an adiabatic process in which a system-coupled parameter $\lambda$ changes from $\lambda_0$ to $\lambda_1$ during time interval $(0, t_S)$ with the system isolated from heat bath,

$<\exp(-\beta W)>$ is given by[11]

$$\langle \exp(-\beta W) \rangle = \frac{\int_{\Omega(t_S)} d\Gamma^* \exp[-\beta H_1(\Gamma^*)]}{\int_{\Omega_0} d\Gamma_0 \exp[-\beta H_0(\Gamma_0)]}. \tag{7}$$

In Eq. (7), $\int_{\Omega_0} d\Gamma_0$ denotes the sum over every microscopic states $\Gamma_0$ of the system with $\lambda = \lambda_0$ at thermal equilibrium, whereas $\int_{\Omega(t_S)} d\Gamma^*$ denotes sum over microscopic states $\Gamma^*(t_S | \Gamma_0)$ of the isolated system at time $t_S$ that has evolved from the initial microscopic states $\Gamma_0$ of the system. Therefore, the sufficient and necessary condition for the validity of Eq. (1) for an adiabatic process in which a system-coupled parameter $\lambda$ changes from $\lambda_0$ to $\lambda_1$ during time interval $(0, t_S)$ is that the phase space extension of the isolated system at time $t_S$, or at the very end of the adiabatic process, should coincide with that of the system with $\lambda = \lambda_1$ in thermal equilibrium with a heat bath to satisfy

$$\int_{\Omega_1} d\Gamma \exp[-\beta H_1(\Gamma)] = \int_{\Omega(t_S)} d\Gamma^* \exp[-\beta H_1(\Gamma^*)]. \tag{8}$$

As a matter of fact, if Eq. (8) is satisfied, Eq. (1) holds for an adiabatic process irrespective of the size of the phase-space extension of an initial system state. However, whenever this condition is not satisfied, Eq. (1) breaks down even when the infinite phase space extension is taken into account, as in the examples of the system we consider here and the adiabatic expansion of a gas into vacuum.

In our model system, the ratio of phase-space extension of thermodynamic state $(R_1, T)$ to that of the initial thermodynamic state $(R_0, T)$ is given by $(R_1/R_0)^2$, and greater than 1 if $R_1$ is larger than $R_0$. Since the phase space extension conserves during an adiabatic expansion process, according to principle of conservation in phase,[15,16] the greater phase space extension of thermodynamic state $(R_1, T)$ can never be covered during an adiabatic process from thermodynamic state $(R_0, T)$ with the smaller phase space extension. For the adiabatic expansion of an ideal gas into vacuum also, it can be shown that the phase space extension of the isolated ideal gas system at time $t_S$ cannot span entire available phase-space extension of the ideal gas, which will be discussed in more detail elsewhere.

In this work, we report the result of an exact theoretical test of Jarzynski relation for a system in an isotropic potential field. The result tells us that Jarzynski relation breaks down for an adiabatic process even when the entire infinite momentum space of the initial system state is taken into account, in contradiction to recent arguments supporting Jarzynski relation.

## Acknowledgement

The author would like to thank Dr. Changbong Hyeon for helpful comments.